\documentclass[a4paper,preprint,aps,pra,showkeys,superscriptaddress]{revtex4-2}

\usepackage{lineno,hyperref,soul,amsmath,amssymb,color,graphicx,bm}

\usepackage{graphicx}
\usepackage[caption=false]{subfig}   
\usepackage{xcolor}       
\usepackage{longtable}
\usepackage{xparse}
\NewExpandableDocumentCommand\mcc{O{1}m}
{\multicolumn{#1}{l}{#2}}
\usepackage[utf8]{inputenc} 
\usepackage[english]{babel}
\usepackage{enumitem, etoolbox, tabularx, makecell, booktabs, float, nccmath}
\newcolumntype{P}[1]{>{\abovedisplayskip=\abovedisplayshortskip \belowdisplayskip=\belowdisplayshortskip}p{#1}}
\usepackage{booktabs,siunitx,makecell}
\usepackage{dcolumn}
\usepackage[colorinlistoftodos]{todonotes}
\usepackage{marginnote}
\usepackage{fancyhdr}
\usepackage{adjustbox}
\usepackage{multirow}
\usepackage[normalem]{ulem}
\usepackage{xcolor}
\usepackage{lipsum}

\makeatletter
\usepackage{miller}
\newcommand\footnoteref[1]{\protected@xdef\@thefnmark{\ref{#1}}\@footnotemark}
\makeatother
 
\reversemarginpar

\hypersetup{
	pdfnewwindow=false,      
	colorlinks=true,       
	linkcolor=red,          
	citecolor=blue,        
	filecolor=magenta,      
	urlcolor=cyan           
}
\usepackage{siunitx}
\usepackage{xfrac}
\usepackage{accents}
\numberwithin{equation}{section}
\makeatletter
\def\tagform@#1{\maketag@@@{(#1)\@@italiccorr}}
\makeatother
\let\orgautoref\autoref

\let\orgautoref\autoref

\let\orgautoref\autoref

\renewcommand{\autoref}[1]
{%
	\def\equationautorefname{Eq.}%
	\def\figureautorefname{Fig.}%
	\def\paperautorefname{paper}%
	\def\subfigureautorefname{Fig.}%
	\orgautoref{#1}%
}

\renewcommand{\autoref}[1]
{%
	\def\equationautorefname{Eq.}%
	\def\figureautorefname{Fig.}%
	\def\paperautorefname{paper}%
	\def\subfigureautorefname{Fig.}%
	\orgautoref{#1}%
}

\makeatletter
\def\tagform@#1{\maketag@@@{\ignorespaces#1\unskip\@@italiccorr}}%
\let\orgtheequation\theequation%
\def\theequation{(\orgtheequation)}%
\makeatother%
\usepackage{soul}
\usepackage{longtable}
\usepackage{enumitem}
\usepackage{etoolbox}
\makeatletter
\patchcmd\frontmatter@PACS@format{\addvspace{11\p@}}{}{}{}
\pretocmd\frontmatter@keys@format{\addvspace{11\p@}}{}{}
\patchcmd{\titleblock@produce}
  {\@pacs@produce\@pacs\@keywords@produce\@keywords}
  {\@keywords@produce\@keywords\@pacs@produce\@pacs}
  {}{}
\makeatother
\newlength\mylength
\usepackage[caption=false]{subfig}
\usepackage{chemformula}
\usepackage{makecell}

\begin{document}
\title{Achieving DFT accuracy in short range ordering and stacking fault energy using moment tensor potential for CoCrFeNi and CoCrNi}
\author{Mashroor S. Nitol}
\email{mash@lanl.gov}
 \affiliation{Center for Integrated Nanotechnologies, Los Alamos National Laboratory, Los Alamos, NM 87544, USA}
 \author{Artur Tamm}
 \affiliation{Institute of Physics, University of Tartu, 50411 Tartu, Estonia} 
  \author{Subah Mubassira}
\affiliation{School of Aerospace and Mechanical Engineering, University of Oklahoma, Norman, OK, 73019-1052, USA}
 \author{Shuozhi Xu}
\affiliation{School of Aerospace and Mechanical Engineering, University of Oklahoma, Norman, OK, 73019-1052, USA}
\author{Saryu J. Fensin}
 \affiliation{Center for Integrated Nanotechnologies, Los Alamos National Laboratory, Los Alamos, NM 87544, USA}

\begin{abstract}
Medium-entropy alloys (MEAs), such as CoCrFeNi and CoCrNi, are regarded as promising structural materials due to their exceptional mechanical and thermal properties, which are governed by complex chemical disorder and atomic-scale interactions. Fundamental insights into these systems have been obtained through density functional theory (DFT), however the scalability limitations of DFT have prevented investigations of large-scale phenomena. Classical interatomic potentials have been used to address this challenge; however, they have been shown to lack the fidelity needed to capture many-body interactions and chemical short-range ordering (CSRO) effects. In the present work, a machine-learned Moment Tensor Potential (MTP) was developed to bridge this gap between accuracy and efficiency. The potential was trained on a comprehensive DFT database spanning unary to quaternary 
configurations and was demonstrated to reproduce energies, forces, and stresses with near-DFT accuracy across diverse structural and chemical environments. Elastic properties were predicted with high fidelity, and compositional trends in bulk and shear moduli were recovered in agreement with DFT. Hybrid Monte Carlo/molecular dynamics simulations were performed, through which CSRO was captured. Key features reported by DFT, including Cr–Cr and Fe–Fe repulsion and Ni–Cr ordering, were reproduced. Stacking fault energetics were also modeled, with ISF energies of approximately 54 mJ/m$^2$ for CoCrNi and 36 mJ/m$^2$ for CoCrFeNi, values consistent with DFT predictions. Local chemical environment effects on stacking faults were further resolved, with Co-rich planes found to reduce, and Cr- or Fe-rich planes found to increase, the stacking fault energy. By enabling large-scale, high-fidelity simulations at a fraction of the computational cost of DFT, the developed MTP establishes a robust framework for predictive modeling of the thermodynamic stability, defect behavior, and mechanical response of FCC MEAs.
\end{abstract}

\keywords{CoCrFeNi, medium entropy alloys, phase transition, molecular dynamics, machine learning, interatomic potential}
\maketitle 
\newpage

\section{Introduction}

Medium-entropy alloys (MEAs) represent a transformative approach to alloy design, characterized by the incorporation of three or four principal elements in near-equiatomic ratios, each typically occupying 5--35 at.\%\cite{yeh2004nanostructured,yeh2007high}. The medium configurational entropy promotes the stabilization of simple solid solution phases, often with face-centered cubic (FCC) or body-centered cubic (BCC) structures, rather than complex intermetallic compounds \cite{laurent2015insights}. Among MEAs, the equiatomic CoCrFeNi and CoCrNi alloys have emerged as model systems due to their exceptional combination of mechanical strength, ductility, thermal stability, and corrosion resistance. These attributes make these alloys promising candidates for demanding applications, including cryogenic structural components, energy systems, and next-generation aerospace materials \cite{bhattacharjee2014microstructure,stepanov2015effect}.

The microstructure of CoCrFeNi and CoCrNi is predominantly a single-phase FCC solid solution with a random distribution of constituent elements \cite{cantor2004microstructural,gludovatz2014fracture}. This phase stability arises from the medium entropy of mixing, which suppresses the formation of intermetallic compounds and promotes a homogenous structure even after prolonged thermal exposure \cite{sathiaraj2015analysis,gludovatz2015processing}. As-cast CoCrFeNi typically exhibits a dendritic morphology, and at finer scales, nanoprecipitates may form as a result of sluggish atomic diffusion inherent to MEAs \cite{liu2013grain,laplanche2016microstructure}. The alloy retains a stable FCC lattice across a broad temperature range \cite{otto2013influences,schuh2015mechanical}, with deformation mechanisms dominated by dislocation slip and nanotwinning, particularly under cryogenic conditions. This microstructural simplicity, combined with complex atomic-scale interactions, underpins its remarkable mechanical performance and makes it an ideal platform for studying fundamental processes in multicomponent metallic systems \cite{senkov2011mechanical,okamoto2016size}.

Extensive first-principles calculations based on density functional theory (DFT) have been conducted to investigate the fundamental properties of CoCrFeNi and CoCrNi \cite{aidhy2024chemical,oh2016lattice,dong2018thermal,liu2018stacking,werner2024experimental,chang2021microstructure,kumar2024cr,fedorov2020phase,moravcik2022impact,wang2018magnetic}. In particular, the energetics of point defects and small vacancy clusters have been evaluated to understand defect stability and migration behavior. These studies have revealed unique characteristics in defect interactions, which are significantly influenced by the complex chemical disorder inherent in the alloy. It was demonstrated that CoCrFeNi and CoCrNi possess an enhanced irradiation resistance due to the difficulty of vacancy cluster formation \cite{xu2022comparative,kang2023density}. Binding energy calculations indicated that di-vacancy clusters are marginally stable, while tri-vacancy clusters are unstable, limiting the growth of larger vacancy clusters and suppressing void swelling \cite{chen2018vacancy,xu2021irradiation}. Furthermore, thermodynamic modeling revealed that phase stability is governed not solely by configurational entropy but also by significant contributions from vibrational, electronic, and magnetic entropies \cite{ma2015ab}. Thermal expansion and bulk modulus predictions showed good agreement with experimental data, confirming the robustness of the first-principles approach \cite{niu2016first}.
Despite these insights, the inherent limitations of DFT in exploring large-scale behavior, such as microstructural evolution under extended irradiation or mechanical loading, remain pronounced. The atomistic models used in DFT are restricted to supercells of a few hundred atoms due to computational expense, thereby limiting the accessible timescales and defect concentrations. 

Classical interatomic potentials, such as the embedded-atom method (EAM) \cite{daw1984embedded}, modified embedded-atom method (MEAM) \cite{lee2000second} and Lennard-Jones (LJ) \cite{jones1924determination} potentials, have been widely applied to investigate the properties of CoCrFeMnNi high-entropy alloys (HEAs). MEAM-based molecular dynamics (MD) simulations successfully captured critical physical metallurgical phenomena, such as sluggish diffusion and microtwinning at cryogenic temperatures \cite{choi2018understanding}. However, the MEAM potential relied heavily on binary parameter fitting and assumed transferability to multicomponent systems, an assumption that introduces inaccuracies, particularly for properties sensitive to local chemical complexity.
In contrast, an LJ-type effective pair potential, developed specifically for the CoCrFeMnNi system, provided an alternative minimalist framework \cite{groger2020effective}. By parameterizing the interaction strength based on equilibrium structures of pure elements and applying regular mixing rules, this approach demonstrated good qualitative agreement with experimental elastic constants, stacking faults, and dislocation behaviors. Nevertheless, the pairwise nature of the LJ potential inherently limits its ability to describe multi-body effects critical for accurately capturing complex defect interactions and CSRO, which are prominent in MEAs and HEAs.
Despite their utility, both MEAM and LJ potentials face fundamental challenges in modeling MEAs and HEAs. First, the parametrization process for classical potentials is often empirical and lacks systematic control over accuracy. Second, their functional forms impose inherent limitations in representing the highly anharmonic and chemically disordered energy landscapes characteristic of multicomponent alloys \cite{wu2024machine,cao2025capturing}. Third, classical potentials struggle to reproduce thermodynamic quantities, such as vibrational entropy or point defect formation energies, at an accuracy level comparable to first-principles methods. Consequently, predictions of properties such as vacancy cluster stability, stacking fault energies, and diffusion barriers remain semi-quantitative at best \cite{yin2021atomistic}.
Recent advances in machine learning (ML)-based interatomic potentials have offered a promising route to overcome these deficiencies \cite{deringer2019machine, men2023understanding,zhou2022thermodynamics,zheng2023multi,wang2024unraveling,nitol2022machine,nitol2023hybrid,zuo2020performance}. Several classes of ML potentials exist, including neural network potentials (e.g., Behler-Parrinello type) \cite{behler2016perspective}, Gaussian approximation potentials (GAP) \cite{bartok2010gaussian}, spectral neighbor analysis potentials (SNAP) \cite{thompson2015spectral}, and moment tensor potentials (MTP) \cite{shapeev2016moment}. In particular, MTP framework has demonstrated the capability to bridge the gap between quantum mechanical accuracy and classical MD efficiency \cite{kwon2023accurate,wang2024efficient,luo2023set,zuo2020performance,yin2021atomistic,nitol2025moment}. MTPs systematically approximate the potential energy surface by fitting a hierarchy of tensorial descriptors to a large database of first-principles calculations \cite{novikov2020mlip}. Importantly, MTPs can naturally capture many-body interactions and local environment dependencies \cite{yin2021atomistic}.

A machine-learning potential \cite{cao2025capturing}(hereafter referred to as MTP$\_$Cao) was recently developed for CoCrNi by training exclusively on ternary CoCrNi configurations, with the objective of capturing CSRO and reproducing key quantities such as the generalized stacking fault energy (GSFE). While this potential demonstrated satisfactory performance for the targeted alloy and properties, its applicability remains confined to the CoCrNi system and does not extend to solute effects or to the thermodynamic and mechanical properties of the constituent unary elements. In the present work, new MTPs for CoCrNi and CoCrFeNi were constructed using a systematically expanded training database incorporating unary, binary, ternary, and quaternary configurations, thereby enhancing both accuracy and transferability. The inclusion of unary structures was found to be critical for properly constraining elemental energetics, improving phase stability predictions, and ensuring reliable extrapolation to dilute solute concentrations and defect configurations. This comprehensive training strategy enables robust modeling of diverse phenomena—including defect energetics, segregation tendencies, and temperature-dependent properties—across a broader compositional and configurational space than previously attainable with a purely ternary-trained potential. Here, the newly developed MTP for the CoCrFeNi and CoCrNi systems brings several advancements over traditional classical potentials. First, it offers near DFT-level accuracy in predicting energies, forces, and stresses, as validated through comparison with extensive DFT datasets. Second, the potential demonstrates excellent transferability across a wide range of configurations, including defective structures, thermalized states, and varying chemical compositions. Third, MTPs enable reliable modeling of subtle effects such as vibrational properties, vacancy migration barriers, and stacking fault energies, which are crucial for understanding deformation and radiation damage phenomena in MEAs.

\section{Result}
\subsection{Force validation}

\begin{figure}[htbp]
    \centering
    \includegraphics[width=\textwidth]{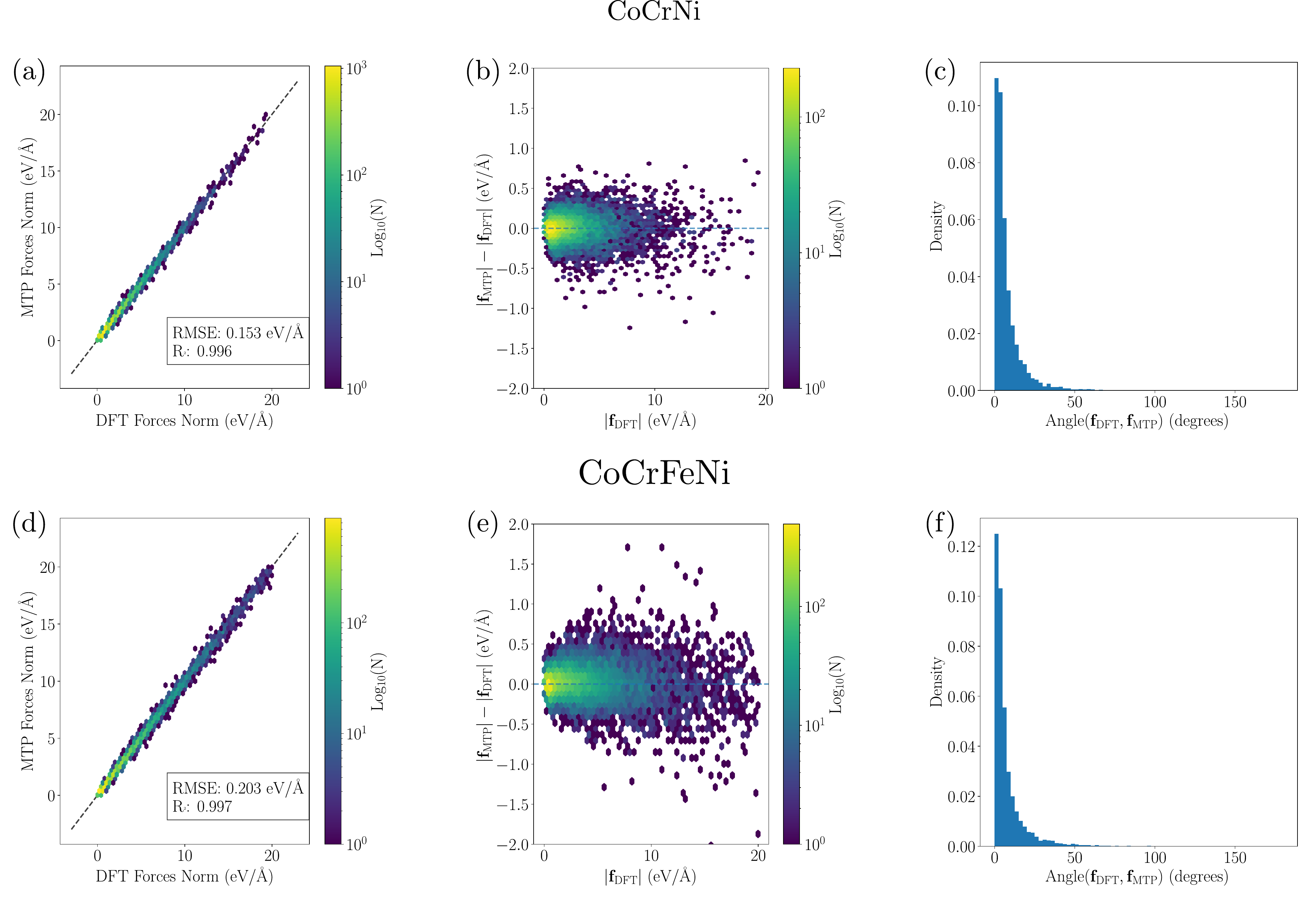}
 \caption{Force validation of the MTP against DFT for CoCrNi (a–c) and CoCrFeNi (d–f). 
(a,d) Parity plots of force norms $|\mathbf f|$ shown as hexbin density maps (color indicates $\log_{10}(N)$) exhibit tight clustering along the diagonal. 
(b,e) The magnitude deviation $\Delta=|\mathbf f_{\mathrm{MTP}}|-|\mathbf f_{\mathrm{DFT}}|$ versus $|\mathbf f_{\mathrm{DFT}}|$ is narrowly centered around zero across the dataset. 
(c,f) Distributions of the angle between $\mathbf f_{\mathrm{DFT}}$ and $\mathbf f_{\mathrm{MTP}}$ are concentrated at small values, indicating strong directional agreement.}

    \label{force_validation}
\end{figure}

The fidelity of the MTP was assessed by comparing predicted atomic forces with DFT references for the ternary CoCrNi and quaternary CoCrFeNi datasets (\autoref{force_validation}). Panels (a) and (d) show parity plots of force norms, with points tightly clustered along the diagonal (color indicates $\log_{10}(N)$), yielding RMSEs of 0.153~eV/\AA\ (CoCrNi) and 0.203~eV/\AA\ (CoCrFeNi) with $R^{2}=0.996$–0.997. Component-wise errors are similarly low: for CoCrNi, RMSE$_{x,y,z}=\{0.146,\,0.145,\,0.145\}$~eV/\AA\ with $R^{2}=0.995$ for each; for CoCrFeNi, RMSE$_{x,y,z}=\{0.186,\,0.184,\,0.184\}$~eV/\AA\ with $R^{2}=0.996$. Panels (b) and (e) plot the difference in magnitudes, $\Delta=|\mathbf f_{\mathrm{MTP}}|-|\mathbf f_{\mathrm{DFT}}|$, versus $|\mathbf f_{\mathrm{DFT}}|$; the distributions are narrowly centered around zero, with mean $|\Delta|=0.112$ (median $0.083$)~eV/\AA\ and mean relative $|\Delta|=8.61\%$ for CoCrNi, and mean $|\Delta|=0.139$ (median $0.095$)~eV/\AA\ and mean relative $|\Delta|=9.17\%$ for CoCrFeNi. Panels (c) and (f) report the angle between the DFT and MTP force vectors; the orientations are strongly aligned, with mean (median) angles of $7.78^\circ$ ($4.54^\circ$) for CoCrNi and $7.82^\circ$ ($4.13^\circ$) for CoCrFeNi, and 95th percentiles of $25.13^\circ$ and $27.20^\circ$, respectively. The close agreement between predicted and reference forces confirms the reliability of the trained MTP in capturing atomic-scale interactions, which is essential for accurate modeling of mechanical behavior, defect energetics, and diffusion mechanisms in these two MEAs.

\subsection{Elastic properties}

To ensure that the atomic structures represent the lowest-energy configurations, a \(10 \times 10 \times 10\) supercell constructed from a four-atom FCC unit cell was used (4000 atoms). First, an equiatomic‐composition supercell was constructed; it followed that hybrid Monte Carlo (MC)/MD simulations \cite{sadigh2012scalable} were performed at 300K, where MC trial swaps explore configurational space while MD relaxations minimize atomic forces and stresses. This approach allows for efficient convergence toward minimum-energy configurations by combining compositional and positional optimization. The hybrid simulations were conducted for 5,000,000 steps, with an energy convergence criterion requiring that the change in total energy after each MC cycle be less than \(1 \times 10^{-6}\)~eV/atom, following the methodology outlined by ~\citet{mubassira2025chemical}. The resulting atomic arrangements are then relaxed into low-energy states, naturally incorporating CSRO effects as a consequence of energy minimization, consistent with previous findings for CoCrFeNi-based alloys \cite{mizuno2022prediction,zhang2020short,chen2021direct}. The elastic constants of CoCrNi and CoCrFeNi obtained from MTP, EAM, MEAM, DFT, and experiment are summarized in Table~\ref{tab elastic_constants}. For both alloys, the lattice constants predicted by all methods lie within the narrow range of 3.53--3.56~\AA, in excellent agreement with experimental measurements, confirming that the FCC structure is well captured across approaches. 

In terms of elastic response, the MTP predictions are in closest agreement with DFT and experimental data. For CoCrNi, MTP yields $C_{11}=245$~GPa, $C_{12}=152$~GPa, and $C_{44}=148$~GPa, values nearly identical to the experimental measurements of $C_{11}=249$~GPa, $C_{12}=156$~GPa, and $C_{44}=142$~GPa. Similarly, for CoCrFeNi, MTP reproduces the balance among $C_{11}$, $C_{12}$, and $C_{44}$ within the DFT range, further confirming its fidelity. In contrast, EAM and MEAM systematically underestimate the shear modulus $C_{44}$ (e.g., $C_{44}=95$~GPa and $82$~GPa in CoCrNi), while slightly overestimating $C_{12}$, leading to an overprediction of the bulk modulus $B$. These deficiencies reflect their empirical nature and lack of explicit electronic structure and magnetic contributions. 

\begin{table}[!htbp]
\centering
\caption{Comparison of elastic constants $C_{ij}$ (GPa) and lattice constant $a$ (\AA) of two equiatomic MEAs obtained from MTP, EAM, MEAM, special quasirandom structures (SQS) from DFT, and experiment. Percent errors in parentheses are relative to DFT.}
\resizebox{\textwidth}{!}{%
\begin{tabular*}{\linewidth}{@{\extracolsep{\fill}}llccccc}
\hline \hline
Alloy & Method & $a$ (\AA) & $C_{11}$ & $C_{12}$ & $C_{44}$  & $B$ \\
\hline
\multirow{5}{*}{CoCrFeNi}
& MTP (SQS)                        & 3.55 ($-0.3\%$) & 214 ($-10.1\%$) & 138 ($-14.8\%$) & 136 ($-21.4\%$) & 164 ($-21.9\%$) \\
& MTP (SRO)                        & 3.55 ($-0.3\%$) & 246 ($+3.4\%$)  & 139 ($-14.2\%$) & 149 ($-13.9\%$) & 175 ($-16.7\%$) \\
& MEAM \cite{choi2018understanding} & 3.54 ($-0.6\%$) & 240 ($+0.8\%$)  & 166 ($+2.5\%$)  & 80 ($-53.8\%$)  & 190 ($-9.5\%$) \\
& DFT  \cite{ge2018effect}          & 3.56      & 238              & 162              & 173              & 210 \\
& Expt \cite{lucas2011magnetic}     &                  & 238              & 151              & 168              & 182 \\
\hline
\multirow{9}{*}{CoCrNi}
& MTP (SQS)                        & 3.54 ($+0.3\%$) & 247 ($-7.5\%$)  & 149 ($-18.1\%$) & 145 ($-18.5\%$) & 182 ($-13.3\%$) \\
& MTP (SRO)                        & 3.53 ($0.0\%$)  & 266 ($-0.4\%$)  & 145 ($-20.3\%$) & 155 ($-12.9\%$) & 186 ($-11.4\%$) \\
& MTP\_Cao (SQS) \cite{cao2025capturing} & 3.52 ($-0.3\%$) & 300 ($+12.4\%$) & 189 ($+3.8\%$)  & 153 ($-14.0\%$) & 226 ($+7.6\%$) \\
& MTP\_Cao (SRO) \cite{cao2025capturing} & 3.52 ($-0.3\%$) & 309 ($+15.7\%$) & 196 ($+7.7\%$)  & 156 ($-12.4\%$) & 234 ($+11.4\%$) \\
& EAM \cite{li2019strengthening}   & 3.56 ($+0.8\%$) & 252 ($-5.6\%$)  & 178 ($-2.2\%$)  & 95 ($-46.6\%$)  & 203 ($-3.3\%$) \\
& MEAM \cite{choi2018understanding} & 3.54 ($+0.3\%$) & 248 ($-7.1\%$)  & 166 ($-8.8\%$)  & 82 ($-53.9\%$)  & 192 ($-8.6\%$) \\
& DFT  \cite{ge2018effect}         & 3.53             & 267              & 182              & 178              & 210 \\
& Expt (298K) \cite{laplanche2020processing} & 3.56 & 249 & 159 & 138 & 189 \\
& Expt (fit) \cite{laplanche2020processing}  &      & 255 & 159 & 146 & 191 \\
\hline \hline
\end{tabular*}
}
\label{tab elastic_constants}
\end{table}


\subsection{Optimal solid‐solution prediction}

To assess the accuracy of MTP in predicting the mechanical properties of chemically disordered CoCrFeNi-based alloys, a series of  MC/MD simulations were performed on non-equiatomic CoCrFeNi compositions. Similarly, elastic moduli were then computed from these relaxed atomic structures and compared to DFT calculations performed using identical compositions and the same total number of configurations \cite{niu2016first}.\\

\begin{figure}[!htbp]
\centering
\includegraphics[width=\textwidth]{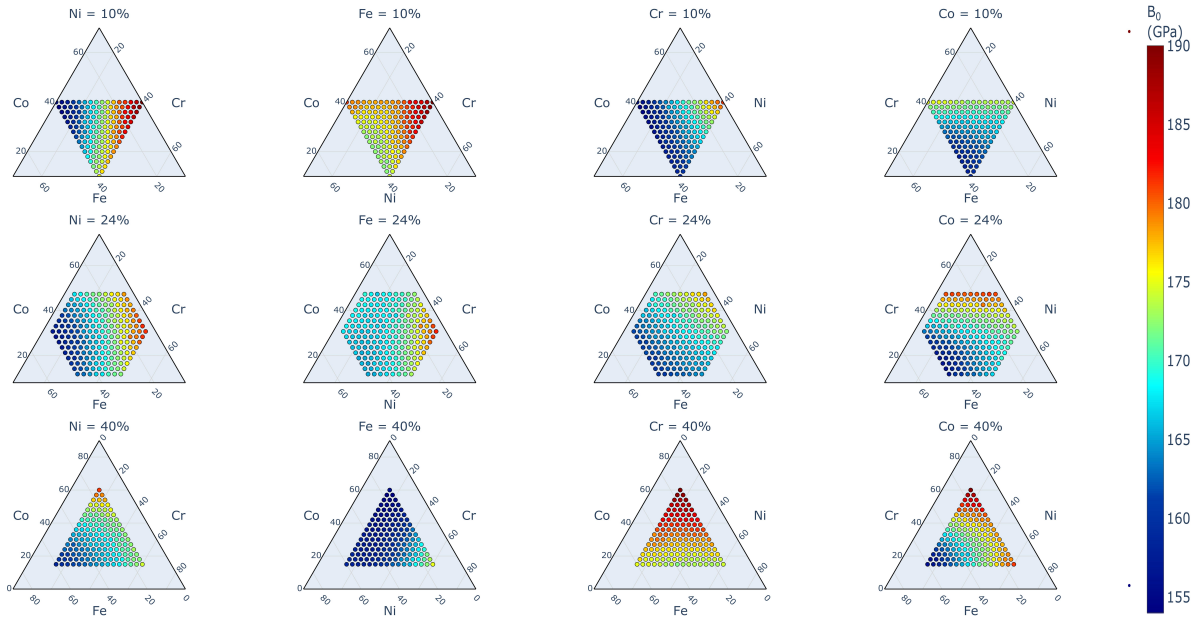}

\caption{Predicted bulk modulus, $B_0$ (GPa), for non-equiatomic CoCrFeNi alloys obtained from MTP after MC/MD relaxation of CSRO configurations. Each cell in the grid corresponds to a unique composition (ordered as \emph{Co, Cr, Fe, Ni}) and is colored by the resulting $B_0$; darker/lighter shades indicate lower/higher stiffness, respectively. Although the MTP was trained only at the equiatomic composition, the map reveals smooth, physically consistent trends across composition space: Fe enrichment is associated with reduced $B_0$ (a low-$B_0$ corridor in Fe-rich regions), while Fe-lean compositions exhibit higher $B_0$. The ordering of the lowest- and highest-$B_0$ compositions matches DFT at the same compositions, as summarized in \autoref{tab:bulk_compare}, supporting the transferability of the model beyond its training point.}

\label{fig:bulk_map}
\end{figure}

\begin{figure}[!htbp]
\centering
\includegraphics[width=\textwidth]{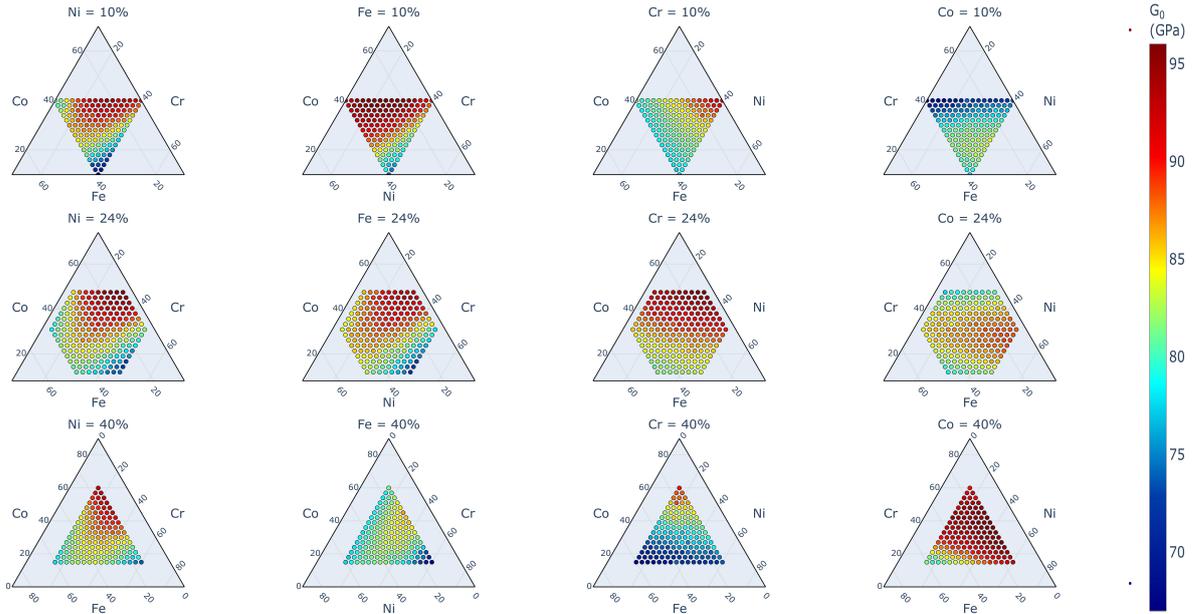}
\caption{Predicted shear modulus, $G_0$ (GPa), for non-equiatomic CoCrFeNi alloys obtained from MTP after MC/MD relaxation of CSRO configurations. The composition grid (ordered as \emph{Co, Cr, Fe, Ni}) is colored by the resulting $G_0$, highlighting systematic trends in rigidity with chemistry. Co enrichment correlates with increased $G_0$ (high-$G_0$ band in Co-rich regions), whereas low-Co/high-Cr combinations tend to yield reduced $G_0$; compositions that are simultaneously Co-rich and Ni-lean are among the stiffest in shear. The MTP reproduces the DFT ranking of the lowest- and highest-$G_0$ compositions (\autoref{tab:shear_compare}), indicating that both models capture the dominant role of Co (and the secondary influence of Cr/Ni) in controlling shear rigidity across this quaternary space.}

\label{fig:shear_map}
\end{figure}

Bulk modulus ($B_0$) and shear modulus ($G_0$) were computed from MTP-relaxed structures and compared with DFT values at the same compositions. \autoref{fig:bulk_map} and \autoref{fig:shear_map} show the predicted distributions of $B_0$ and $G_0$, respectively, across a representative set of non-equiatomic compositions. Although the MTP was trained only on equiatomic CoCrFeNi, it reproduces the DFT-predicted ordering of the compositions with the lowest and highest elastic moduli. Consistent with DFT trends, Fe enrichment is associated with the lowest $B_0$, whereas Co enrichment is associated with the highest $G_0$. Concretely, in our top three low-$B_0$ compositions, Fe = 40\% in all cases, while in the top three high-$B_0$ compositions Fe = 10\%. For $G_0$, the three lowest-$G_0$ alloys all contain Co = 10\% and Cr = 40\%, whereas the three highest-$G_0$ alloys have Co = 40\% and Ni = 10\%. Overall, the MTP and DFT results are in qualitative agreement across composition space, capturing the dominant influence of Fe on $B_0$ and Co on $G_0$.As summarized in Tables~\ref{tab:bulk_compare} and~\ref{tab:shear_compare}, the maximum deviation between MTP and DFT results remains within approximately 10--15~GPa for both $B_0$ and $G_0$, indicating excellent quantitative agreement. \\


Furthermore, the trends observed here corroborate earlier first-principles investigations using the EMTO-CPA method~\cite{niu2016first}, which reported that Fe predominantly reduces $B_0$, while Co enhances $G_0$, with Cr exerting a secondary influence through its control over the $B_{0}/G_{0}$ 
ratio. The present results thus demonstrate that the MTP, although trained only on a limited equal-molar composition
generalizes robustly to a broader compositional space and faithfully captures the underlying physical trends dictated by electronic structure effects. This predictive capability highlights the utility of MTPs for efficiently screening mechanical properties across multicomponent alloy systems beyond the explicit limits of the training database.

\begin{table}[!htbp]
\centering
\caption{Top 3 alloys (Co, Cr, Fe, Ni~\%) with the lowest and highest bulk modulus values ($B_{0}$ in GPa) as calculated by MTP and DFT. The MTP predictions show good agreement with DFT results, confirming that Fe has the most significant influence on $B_{0}$: increasing Fe content (e.g., 10\% $\to$ 40\%) corresponds to a decrease in $B_{0}$.}
\label{tab:bulk_compare}
\resizebox{\textwidth}{!}{%
\begin{tabular*} {\linewidth}{@{\extracolsep{\fill}}ll|ll||ll|ll}
\hline \hline
\multicolumn{4}{c||}{Low $B_0$} & \multicolumn{4}{c}{High $B_0$} \\
\hline
MTP  & $B_0$ & DFT  & $B_0$ & MTP  & $B_0$ & DFT  & $B_0$\\
\hline
38, 12, 40, 10 & 154 & 10, 26, 40, 24 & 177 & 40, 40, 10, 10 & 190 & 40, 40, 10, 10 & 232 \\
38, 10, 40, 12 & 154 & 34, 16, 40, 10 & 177 & 38, 40, 10, 12 & 189 & 38, 40, 10, 12 & 231 \\
36, 12, 40, 12 & 155 & 12, 26, 40, 22 & 178 & 38, 40, 10, 12 & 188 & 40, 38, 10, 12 & 231 \\
\hline \hline
\end{tabular*}
}
\end{table}

\begin{table}[!htbp]
\centering
\caption{Top 3 alloys (Co, Cr, Fe, Ni~\%) with the lowest and highest shear modulus values ($G_{0}$ in GPa) as calculated by MTP and DFT. The MTP predictions broadly track DFT trends; increasing Co content corresponds to an increase in $G_{0}$.}
\label{tab:shear_compare}
\resizebox{\textwidth}{!}{%
\begin{tabular*} {\linewidth}{@{\extracolsep{\fill}}ll|ll||ll|ll}
\hline \hline
\multicolumn{4}{c||}{Low $G_0$} & \multicolumn{4}{c}{High $G_0$} \\
\hline
MTP  & $G_0$ & DFT  & $G_0$ & MTP  & $G_0$ & DFT  & $G_0$\\
\hline
10, 40, 40, 10 & 67 & 10, 40, 16, 34 & 84 & 40, 28, 10, 22 & 96 & 40, 16, 34, 10 & 117 \\
10, 40, 34, 16 & 68 & 10, 40, 18, 32 & 84 & 40, 24, 10, 26 & 96 & 40, 14, 36, 10 & 117 \\
10, 40, 36, 14 & 68 & 10, 40, 14, 36 & 84 & 40, 18, 10, 32 & 96 & 40, 18, 32, 10 & 117 \\
\hline \hline
\end{tabular*}
}
\end{table}

\subsection{Short range ordering}

CSRO in MEAs is increasingly recognized as a critical structural feature that governs their thermodynamic stability and mechanical performance \cite{schweika1988neutron,schonfeld1988short,chen2021direct}. Contrary to the idealized view of MEAs as random solid solutions, recent computational and experimental evidence highlights that local chemical ordering, particularly at low temperatures, can significantly affect stacking fault energies, magnetic configurations, lattice distortions, and melting points \cite{zhang2020short}. CSRO introduces preferential pairwise interactions between atoms, which modify the local atomic environments and reduce configurational entropy --- ultimately stabilizing certain atomic configurations. In alloys like CoCrNi and CoCrFeNi, this ordering leads to the formation of chemically distinct domains, enhancing mechanical strength and resistance to deformation by increasing energy barriers for dislocation motion. Moreover, CSRO can influence the electronic density of states and magnetic interactions by modifying the overlap between atomic orbitals and altering the alignment of local magnetic moments, especially in elements like Cr that exhibit antiferromagnetic tendencies \cite{tamm2015atomic}

In the DFT-informed MC simulations by ~\citet{tamm2015atomic}, strong CSRO was observed in both CoCrNi and CoCrFeNi alloys. Cr–Cr and Fe–Fe pairings exhibited strong repulsion (positive Warren-Cowley parameters), while Ni–Cr and Ni–Fe demonstrated attractive ordering (negative values), indicating a strong preference for heterogeneous atomic pairs. This chemical ordering led to a significant reduction in formation energy and configurational entropy, reinforcing its thermodynamic significance. Notably, magnetic frustration emerged as a key consequence: the Cr sublattice exhibited antiferromagnetic ordering, influenced by the CSRO patterns, whereas other elements remained weakly magnetic. In contrast, results obtained using classical EAM potentials \cite{li2019strengthening}, analyzed by ~\citet{mubassira2025chemical}, showed qualitatively similar but quantitatively weaker CSRO trends. The CoCrNi system modeled with an EAM potential failed to capture the full extent of Cr–Cr repulsion and Ni–Cr affinity seen in DFT, largely due to the limitations of EAM in describing directional bonding and magnetic interactions. EAM relies on parameterized, empirical descriptions of atomic interactions, and lacks the quantum mechanical rigor needed to account for electronic structure effects and magnetism, explaining its reduced accuracy in predicting CSRO magnitudes and their consequences.

Although the MTP is not explicitly trained on spin moments, it is fitted to energy, force, and stress data from spin-polarized DFT calculations. Thus, it implicitly retains magnetic effects while remaining computationally efficient. To model CSRO, hybrid MD/MC simulations are performed in an isothermal–isobaric (NPT) ensemble at 500~K until the system's energy converges. The Warren–Cowley parameter is used to quantify CSRO:
\[
\alpha_{ij}^{n} = \frac{p_{ij}^{n} - c_j}{\delta_{ij} - c_j}
\]
where $n$ is the $n$th-nearest-neighbor shell of a central atom of type $i$, $p_{ij}^{n}$ the probability of finding a $j$-type atom in that shell, $c_j$ the concentration of element $j$, and $\delta_{ij}$ the Kronecker delta. In a random alloy, $\alpha_{ij}^{n} \approx 0$; positive values indicate like-atom segregation ($i=j$), while negative values imply ordering between unlike species ($i \neq j$). The FCC simulation cell used contains 2,916 atoms 

The MTP
 simulations reproduce the overall CSRO trends reported by DFT, though with reduced magnitudes. As summarized in \autoref{sro table}, the qualitative \emph{sign structure} of the first–nearest–neighbor CSRO is reproduced most faithfully by the MTP, whereas the empirical models struggle. For CoCrNi, DFT indicates a pronounced Ni–Cr \emph{attraction} ($\alpha_{\mathrm{Ni\text{–}Cr}}<0$) together with strong \emph{clustering} on Cr–Cr ($\alpha_{\mathrm{Cr\text{–}Cr}}>0$). The EAM potential results from \cite{mubassira2025chemical} shows that it fails on both counts—reversing the Ni–Cr sign and predicting Cr–Cr to be mildly repulsive—although it does capture Cr–Co ordering (with an exaggerated magnitude). MEAM performs less consistently, flipping several key signs (e.g., Ni–Co, Cr–Cr, Cr–Co). By contrast, the MTP recovers the correct signs for all pairs and largely preserves the rank ordering of tendencies; its magnitudes are generally damped relative to DFT, which is typical for models trained to balance accuracy across diverse configurations.

For CoCrFeNi, discrepancies are more pronounced for MEAM: it inverts the DFT signs for most pairs (notably Ni–Fe, Cr–Cr, Cr–Co, Cr–Fe, Co–Co, Fe–Fe), implying a qualitatively different chemical affinity landscape. The MTP, trained on DFT–MC trajectories, again matches the DFT \emph{signs} across all pairs and follows the relative trends (e.g., positive clustering on Ni–Ni, Cr–Cr, Co–Co; ordering on Ni–Cr, Ni–Fe, Cr–Co, Cr–Fe, Co–Fe), even if the absolute magnitudes are systematically smaller. Taken together, these comparisons suggest that the MTP provides a reliable qualitative description of chemical short–range order in both ternary and quaternary spaces, suitable for trend analysis and compositional screening, whereas EAM/MEAM should be applied with caution for CSRO-sensitive predictions.

\begin{table}[!htbp]
\centering
\caption{First–nearest–neighbor WC short-range order parameters $\alpha_{ij}$ (dimensionless) for equiatomic CoCrNi (top block: DFT, EAM, MEAM, MTP) and CoCrFeNi (bottom block: DFT, MEAM, MTP) at 500K. Positive $\alpha_{ij}$ indicates like-species clustering, while negative values indicate chemical ordering between unlike species.}
\label{sro table}
\resizebox{\textwidth}{!}{%
\begin{minipage}{\textwidth}

\begin{tabular*}{\linewidth}{@{\extracolsep{\fill}}lllll}
\hline
\multicolumn{5}{c}{CoCrNi}\\ \hline
Pair  & DFT\cite{tamm2015atomic} & EAM\cite{li2019strengthening} & MEAM\cite{choi2018understanding} & MTP \\
Ni-Ni & 0.12 & 0.47 & 0.04 & 0.01 \\
Ni-Cr & -0.27 & 0.38 & -0.03 & -0.20 \\
Ni-Co & 0.15 & 0.57 & -0.05 & 0.16 \\
Cr-Cr & 0.42 & -0.11 & -0.21 & 0.24 \\
Cr-Co & -0.16 & -0.57 & 0.45 & -0.21 \\
Co-Co & 0.01 & -0.08 & -0.20 & 0.02 \\
\hline
\end{tabular*}

\vspace{0.6em}

\begin{tabular*}{\linewidth}{@{\extracolsep{\fill}}llll}
\hline
\multicolumn{4}{c}{CoCrFeNi} \\ \hline
Pair  & DFT\cite{tamm2015atomic} & MEAM\cite{choi2018understanding} & MTP \\
Ni-Ni & 0.30 & 0.11 & 0.06 \\
Ni-Cr & -0.18 & -0.08 & -0.26 \\
Ni-Co & 0.14 & 0.09 & 0.34 \\
Ni-Fe & -0.27 & 0.11 & -0.24 \\
Cr-Cr & 0.59 & -0.28 & 0.26 \\
Cr-Co & -0.31 & 0.26 & -0.23 \\
Cr-Fe & -0.10 & 0.33 & -0.22 \\
Co-Co & 0.33 & -0.02 & 0.03 \\
Co-Fe & -0.16 & -0.31 & -0.21 \\
Fe-Fe & 0.54 & -0.09 & 0.22 \\
\hline
\end{tabular*}

\end{minipage}
}
\end{table}
The apparent reduction in CSRO magnitudes from MTP relative to DFT can plausibly arise from differences in system size and sampling. In our workflow, MTP permits substantially larger supercells (thousands of atoms) and many more MD steps between MC swaps, which in turn sample a wider spectrum of local environments, long-wavelength vibrational modes, and vibrational entropy contributions. By contrast, DFT calculations are necessarily constrained to smaller supercells and shorter trajectories, which can bias the statistics toward more pronounced CSRO values due to finite-size effects. Accordingly, the discrepancies in $\alpha_{ij}$ magnitudes may reflect ensemble and size differences rather than model deficiency. A systematic, like-for-like comparison—matching supercell sizes, MC/MD schedules, and thermostats—will be required to isolate these effects with precision, and is left for future work.

\begin{figure}[!htbp]
    \centering
    \includegraphics[width=\textwidth]{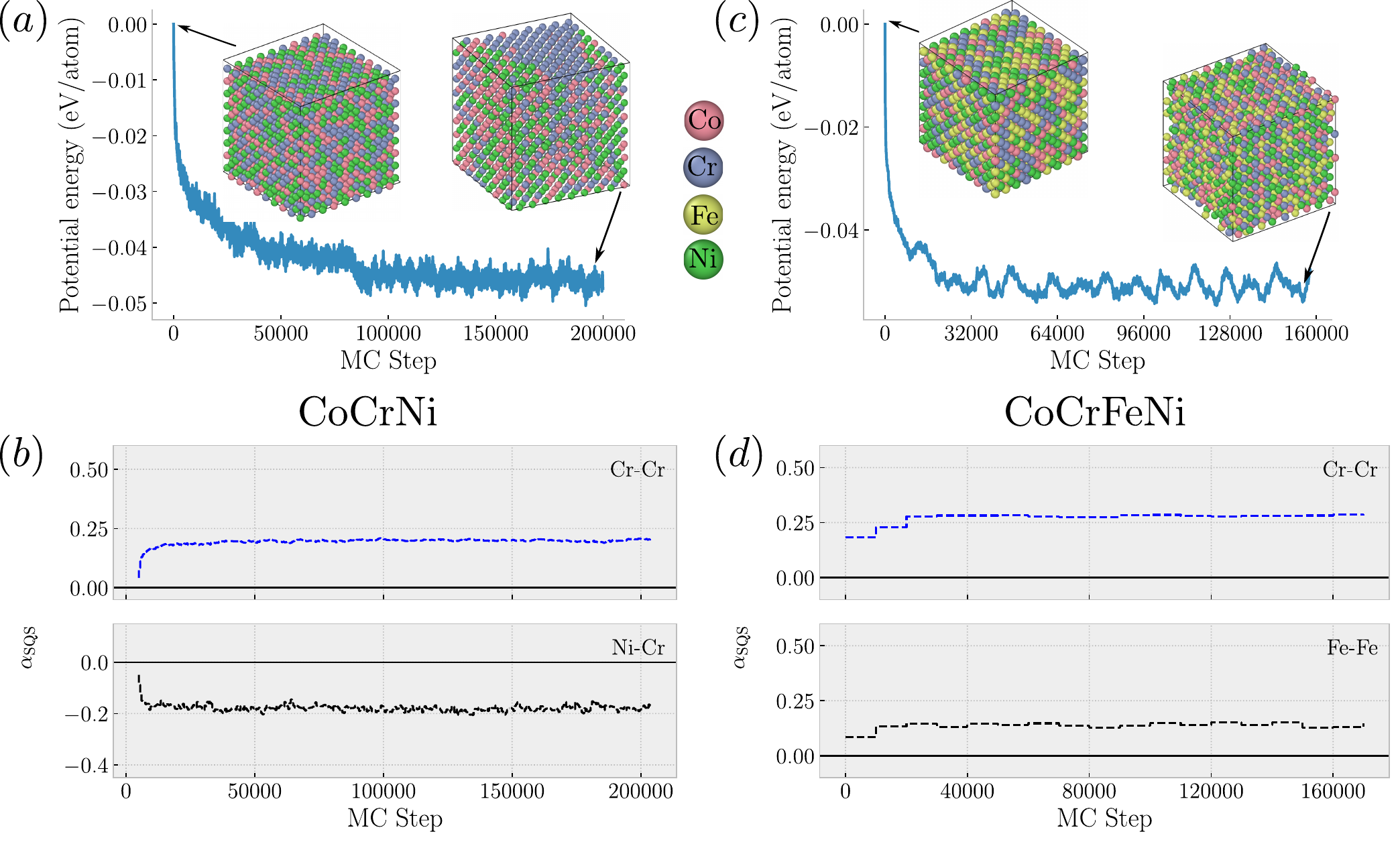}
    \caption{Evolution of potential energy and CSRO during lattice MC simulations at 500~K. (a,c) Potential energy convergence in CoCrNi and CoCrFeNi, respectively. (b,d) Warren–Cowley CSRO parameters in the first-nearest-neighbor shell: (b) Cr–Cr and Ni–Cr in CoCrNi, (d) Cr–Cr and Fe–Fe in CoCrFeNi. 
    }
    \label{sro1}
\end{figure}

\autoref{sro1} shows that in CoCrNi, Cr–Cr pairs exhibit positive $\alpha$ (repulsion), while Ni–Cr pairs show negative $\alpha$ (ordering), consistent with DFT predictions. For CoCrFeNi, both Cr–Cr and Fe–Fe interactions are repulsive, while Ni–Fe and Cr–Ni favor mixing. Compared with DFT, MTP underestimates the strength of like-atom repulsion, reflecting the absence of explicit spin-polarization and electronic structure fidelity. Nonetheless, it captures the qualitative chemical tendencies and provides reliable insight into CSRO evolution at large scales.

\begin{figure}[!htbp]
    \centering
    \includegraphics[width=\textwidth]{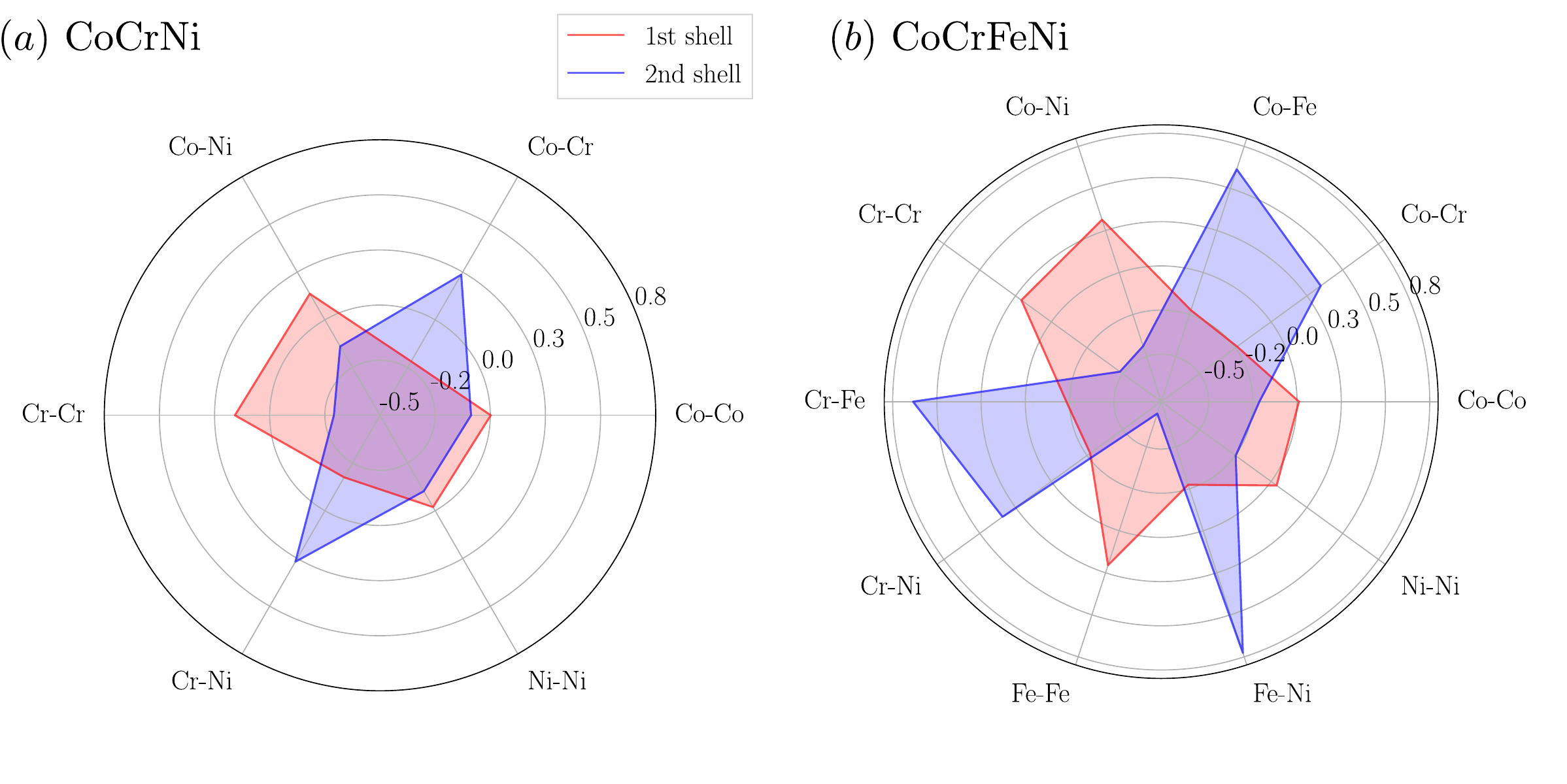}
    \caption{Warren–Cowley CSRO parameters for first- and second-nearest-neighbor shells in (a) CoCrNi and (b) CoCrFeNi, obtained at the end of MTP-based hybrid MC/MD simulations at 500~K. Red regions correspond to the first shell, and blue regions the second shell. Positive values indicate like-atom repulsion, negative values indicate ordering between unlike species. 
    }
    \label{sro2}
\end{figure}

\autoref{sro2} expands the analysis to multiple pairs and shells. In CoCrNi (\autoref{sro2}a), MTP predicts positive $\alpha$ for like-atom interactions (Cr–Cr, Ni–Ni) and negative $\alpha$ for dissimilar pairs (Cr–Ni, Co–Cr), while Co–Co and Co–Ni remain near random, consistent with DFT. For CoCrFeNi (\autoref{sro2}b), the same pattern emerges: strong repulsion for Cr–Cr and Fe–Fe, ordering for Cr–Ni and Fe–Ni, and weak interactions for solvent-like pairs. Importantly, MTP correctly captures the relative trends of CSRO across the first shell, even if it underestimates absolute magnitudes. 

\subsection{Stacking fault energy}

The GSFE is a critical intrinsic property influencing dislocation behavior, phase transformation, and mechanical performance in FCC metals. GSFE values were computed for CoCrNi-based systems using MTP and compared with DFT predictions obtained from SQS. Both random and CSRO configurations were considered within the MTP framework. Simulations were conducted by rigidly shifting the upper half of a supercell along the \{111\} glide plane in the $\langle 112 \rangle$ direction for CoCrFeNi and CoCrNi, generating discrete GSFE curves from 16 unique glide planes to ensure statistical reliability. The GSFE landscape was resolved by calculating both the unstable stacking fault energy ($\gamma_{\mathrm{usf}}$), corresponding to the peak barrier for dislocation nucleation, and the intrinsic stacking fault (ISF) energy ($\gamma_{\mathrm{isf}}$), reflecting post-slip stability. For CSRO structures, configurations were pre-equilibrated via hybrid MD/MC annealing under a variance-constrained semi-grand canonical ensemble. The energy profiles were normalized with respect to slip displacement, and GSFE curves were constructed by sampling displacements over one Burgers vector periodicity. 

Comparative analysis of MTP and DFT results indicates that CSRO enhances both $\gamma_{\mathrm{usf}}$ and $\gamma_{\mathrm{isf}}$, suggesting increased resistance to dislocation motion and stabilization of the FCC phase, respectively. This trend underscores the significant role of local chemical environments in tailoring deformation mechanisms in complex alloys.

\begin{figure}[!htbp]
    \centering
    \includegraphics[width=\textwidth]{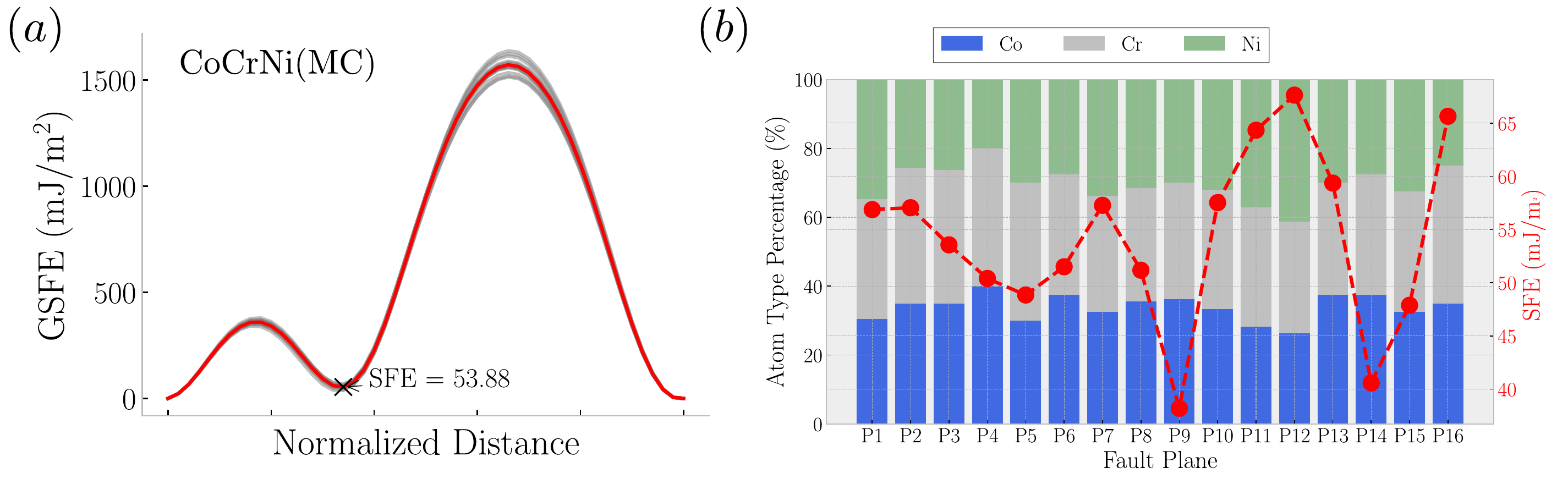}
    \caption{Comparison of stacking fault energetics in CoCrNi obtained from the MC simulation using MTP.  
(a) Average GSFE curve along the $\langle 112 \rangle /6$ slip path, showing an ISF energy of $\gamma_{\mathrm{isf}} \approx 53.9~\mathrm{mJ/m^{2}}$.  
(b) Variation of $\gamma_{\mathrm{isf}}$ across different fault planes and the corresponding local atomic composition}.
    \label{cocrni gsfe}
\end{figure}

\autoref{cocrni gsfe} shows the GSFE obtained from the MTP demonstrates good qualitative agreement with DFT predictions, but underestimates both the magnitude and variability of the ISF energy for CoCrNi. In DFT combined with MC simulations, CoCrNi exhibits an average $\gamma_{\mathrm{isf}} \approx 80~\mathrm{mJ/m^{2}}$ \cite{zhu2023effects}, with a broad distribution ranging from slightly negative values up to $\sim 288~\mathrm{mJ/m^{2}}$, strongly dependent on local chemical environments. MTP predicts a lower average value of $\gamma_{\mathrm{isf}} \approx 54~\mathrm{mJ/m^{2}}$, with a narrower spread of $\sim 40$--$65~\mathrm{mJ/m^{2}}$ across different fault planes. Despite this quantitative discrepancy, the MTP correctly captures the chemistry-dependent trends, where Co-rich fault planes exhibit reduced $\gamma_{\mathrm{isf}}$ while Cr-enriched planes elevate $\gamma_{\mathrm{isf}}$, consistent with DFT results. 

\begin{figure}[!htbp]
    \centering
    \includegraphics[width=\textwidth]{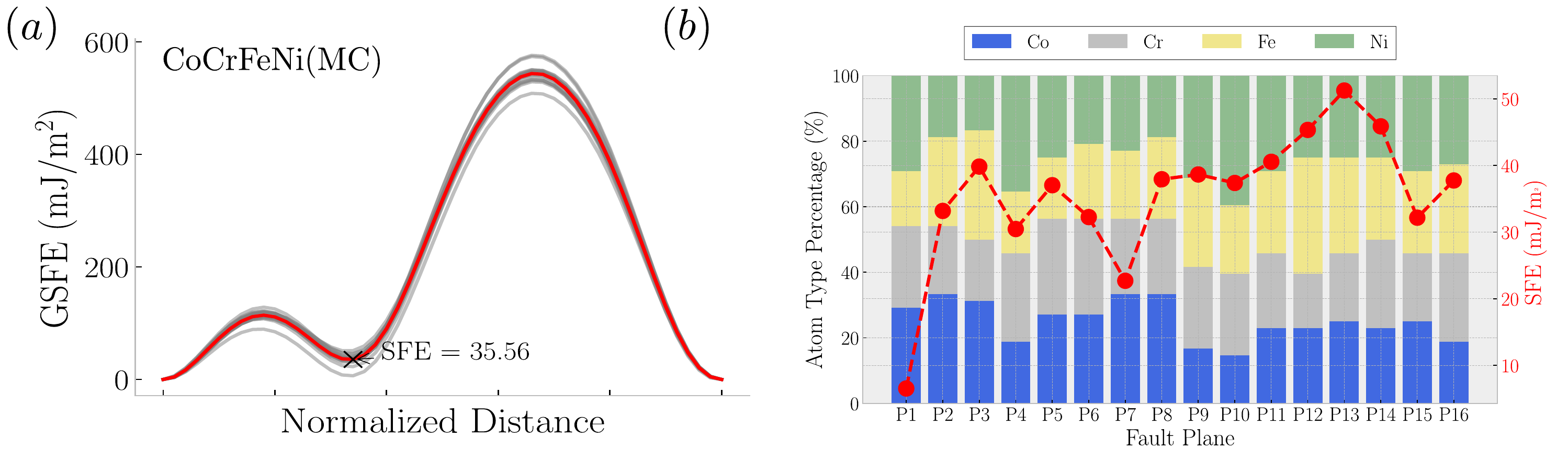}
    \caption{GSFE of CoCrFeNi obtained from the MTP with CSRO configurations.  
(a) Average GSFE curve along the $\langle 112 \rangle /6$ slip path, yielding an ISF energy of $\gamma_{\mathrm{isf}} \approx 35.6~\mathrm{mJ/m^{2}}$.  
(b) Plane-resolved variation of $\gamma_{\mathrm{isf}}$ across different fault planes along with the corresponding local atomic composition, showing clear chemistry-dependent fluctuations. 
}
    \label{cocrfeni gsfe}
\end{figure}

The MTP provides a quantitatively reliable and physically transparent description of stacking fault energetics in CoCrFeNi as well.  
While DFT calculations with MC sampling predict an ISF energy in the range of $17$--$34~\mathrm{mJ/m^{2}}$~\cite{zaddach2013mechanical}, the present MTP with CSRO configurations yields a value of $\gamma_{\mathrm{isf}} \approx 35.6~\mathrm{mJ/m^{2}}$, in excellent agreement with the upper bound of the DFT range.  
Unlike DFT, which is limited to a small number of local configurations, the MTP efficiently samples a wide spectrum of fault-plane chemistries and reveals that Fe- and Ni-rich planes favor higher $\gamma_{\mathrm{isf}}$, whereas Co-rich environments suppress it.  
This chemical sensitivity, illustrated in \autoref{cocrfeni gsfe}, is consistent with first-principles trends but accessible here on significantly larger length scales.  
Moreover, as summarized in ~\autoref{gsfe_comparison}, the MTP  reproduces the relative ordering of GSFE among CoCrFeNi and CoCrNi.  These results underscore that the MTP framework offers a robust and computationally efficient pathway to probe the role of ISFs and CSRO in dictating plasticity in complex concentrated alloys.

\begin{table}[!htbp]
\centering
\caption{Intrinsic stacking fault energies $\gamma_{\mathrm{ISF}}$ (mJ\,m$^{-2}$) for CoCrFeNi and CoCrNi, comparing DFT (SQS), empirical potentials (EAM/MEAM) for CSRO configurations where available, and MTP (CSRO). MTP values are reported as mean~$\pm$~standard deviation across sampled CSRO structures.}
\label{gsfe_comparison}
\resizebox{\textwidth}{!}{%
\begin{tabular*}{\linewidth}{@{\extracolsep{\fill}}lcccc}
\hline\hline
System     & DFT (SQS)                          & EAM (CSRO)\cite{li2019strengthening} & MEAM (CSRO)\cite{choi2018understanding} & MTP (CSRO) \\
\hline
CoCrFeNi   & 17--34 \cite{zaddach2013mechanical} & ---                                   & 88.97$\pm$8                                   & 35.56$\pm$12 \\
CoCrNi     & 80 \cite{zhu2023effects}            & 72.56$\pm$8                                 & 13.82$\pm$6                                     & 53.88$\pm$5 \\
\hline\hline
\end{tabular*}
}
\end{table}


MTP models, when trained carefully on diverse datasets including chemically perturbed structures, can qualitatively reproduce the sensitivity of GSFE to local chemical environment, a feature traditionally absent in empirical potentials such as MEAM~\cite{choi2018understanding} or LJ-type effective potentials~\cite{groger2020effective}. 

\section{Conclusion}

In this work, an MTP was developed for the CoCrFeNi and CoCrNi MEAs and benchmarked against DFT and experimental data. The key findings can be summarized as follows:

\begin{itemize}[leftmargin=*]

\item The MTP quantitatively reproduces the lattice constants
and elastic constants of both CoCrNi and CoCrFeNi, achieving near-DFT accuracy and excellent agreement with experimental measurements. In contrast, classical empirical potentials such as EAM and MEAM systematically underestimate $C_{44}$ 
and overestimate $C_{12}$, resulting in deviations in bulk modulus predictions. The superior accuracy of MTP highlights its ability to capture many-body interactions and chemical disorder effects absent in classical potentials.

\item Although trained on equiatomic CoCrFeNi datasets, the MTP demonstrates robust predictive capability for non-equiatomic compositions. MC/MD simulations using MTP reproduce DFT-predicted trends in bulk and shear moduli across a wide compositional space. Specifically, alloys with higher Fe content exhibit lower bulk moduli, while Co enrichment enhances shear modulus. These trends agree with DFT
predictions, underscoring the MTP's reliability in capturing composition-dependent elastic behavior.

\item The MTP simulations capture the essential features of CSRO in CoCrNi and CoCrFeNi. Consistent with DFT-informed MC studies, MTP predicts strong repulsion between like-species pairs such as Cr–Cr and Fe–Fe, and favorable ordering for dissimilar pairs such as Ni–Cr and Ni–Fe. While the magnitude of CSRO is underestimated relative to DFT due to the absence of explicit spin polarization in MTP, the relative chemical trends are faithfully reproduced. This capability demonstrates that MTP can capture the influence of CSRO on alloy energetics and stability, an essential feature for modeling thermodynamic and mechanical behavior at realistic length scales.

\item The GSFE was systematically evaluated using MTP for both random and CSRO configurations. For CoCrFeNi, MTP predicts an ISF energy of $\sim 35.6~\mathrm{mJ/m^{2}}$, which falls within the upper bound of DFT predictions ($17$--$34~\mathrm{mJ/m^{2}}$). For CoCrNi, MTP yields an ISF energy of $\sim 53.9~\mathrm{mJ/m^{2}}$, lower than the DFT value of $\sim 80~\mathrm{mJ/m^{2}}$ but consistent with the chemical trends reported by first-principles studies. Importantly, MTP captures the sensitivity of ISF to local atomic chemistry, predicting that Co-rich planes reduce $\gamma_{\mathrm{isf}}$ whereas Cr- and Fe-enriched planes increase it. This chemically resolved description of GSFE is not achievable with conventional empirical potentials.

\end{itemize}

The present MTP establishes a robust and transferable framework for the atomistic modeling of CoCrFeNi- and CoCrNi-based MEAs. It effectively reconciles the accuracy of first-principles methodologies with the computational efficiency of empirical potentials, thereby enabling predictive simulations at scales inaccessible to \emph{ab initio} approaches. The potential demonstrates reliability in reproducing elastic properties, CSRO, and stacking fault energetics, which are central to understanding the deformation behavior of these alloys. Collectively, the results reinforce the emerging role of ML interatomic potentials as indispensable tools for the quantitative exploration of defect processes, thermodynamic stability, and mechanical response in complex concentrated alloys.
\section{Methods}
\subsection {DFT training database}

In order to develop a transferable and accurate ML interatomic potential for the CoCrFeNi system, an extensive DFT database was constructed. The database was systematically designed to span a wide range of atomic environments, including unary, binary, ternary, quaternary 
compositions, while capturing both harmonic and anharmonic atomic behaviors across diverse structural and chemical configurations.

For the elemental constituents Co, Cr, Fe, and Ni, a comprehensive set of relaxed crystalline structures was generated. The structural prototypes considered included FCC, BCC, hexagonal close-packed (HCP), simple cubic (SC), $\omega$ phase, A15, diamond cubic, and $\beta$-Sn. Each structure was fully relaxed to obtain equilibrium configurations. To incorporate anharmonic effects and thermal fluctuations, perturbations were introduced around equilibrium atomic positions following the methodology of ~\citet{nitol2023hybrid}. Small random displacements were applied to emulate vibrational motion at finite temperatures, thereby improving the transferability of the trained potential across both harmonic and anharmonic regimes. In addition, surface structures were generated to capture low-coordination atomic environments. Symmetric, non-polar slab models were systematically constructed for each phase using the Pymatgen library~\cite{ong2013python}, with Miller indices enumerated up to a predefined maximum value. Elastic properties were also included by applying finite distortions to the relaxed bulk structures with the \texttt{elastic\_vasp} package~\cite{Elastic_vasp,MISHRA2022111667}. Stress tensors were obtained via DFT, and the independent elastic constants were extracted according to crystal symmetry, thereby embedding mechanical response information in the dataset. To ensure accurate cohesive energies, isolated atom calculations were also performed for each constituent element.

Binary intermetallic structures were then generated for all unique elemental pairs. Although no stable intermetallics exist in the CoCrFeNi system, a wide range of prototype binary structures was considered in order to constrain the ML potential to reproduce the positive mixing enthalpies correctly. These prototypes included B2, Ba, B19 and its derivative B19$'$, B20, B32, B33, B81, L1$_0$, Al$_3$Os$_2$, B82, C1, C14, C22, C38, C40, Cmcm, D1$_a$, D5$_a$, DO$_{19}$, DO$_{22}$, DO$_{23}$, DO$_{24}$,
 and L1$_2$. Unless otherwise required by prototype stoichiometry, equiatomic compositions were adopted. To capture finite-temperature effects, $3 \times 3 \times 3$ supercells of low-energy phases were constructed, including FCC and HCP Co, BCC Cr and Fe, and FCC Ni. Solute atoms were then substituted at concentrations ranging from 1 to 50 at.\% in increments of 5\%. Each configuration was thermally perturbed by random displacements up to 0.6\AA, following the previous work by ~\citet{nitol2025moment}. To further enrich the sampling, isotropic strains of $\pm 5$\% were applied, and for each strain state, 25 distinct thermally perturbed configurations were generated. In this way, solute-host systems were comprehensively explored, including Co-rich (Co–Cr, Co–Fe, Co–Ni in FCC/HCP Co), Cr-rich (Cr–Co, Cr–Fe, Cr–Ni in BCC Cr), Fe-rich (Fe–Co, Fe–Cr, Fe–Ni in BCC Fe), and Ni-rich (Ni–Co, Ni–Cr, Ni–Fe in FCC Ni) environments. This approach ensured robust performance across a wide range of binary alloy compositions and atomic environments.

To further enhance the interpolation capability across multicomponent systems, ternary equiatomic structures were generated. For instance, the Co–Cr–Ni ternary system was constructed using representative phases of its constituents, namely HCP and FCC Co, BCC Cr, and FCC Ni. These structures were relaxed and subsequently perturbed in an analogous manner to the binary dataset in order to emulate thermal fluctuations. Finally, quaternary datasets were prepared for the CoCrFeNi alloy itself, with relaxed FCC structures serving as the basis for additional thermal perturbations.

All DFT calculations were carried out using the Vienna \emph{Ab initio} Simulation Package (VASP)~\cite{hafner2008ab}, version 6.3.2, using the Perdew–Burke–Ernzerhof generalized gradient approximation (PBE–GGA) for exchange-correlation~\cite{perdew1996generalized}. A plane-wave energy cutoff of 520 eV was used, with electronic self-consistency achieved to within $10^{-6}$ eV. Partial occupancies were treated with Gaussian smearing of width 0.2 eV, and Brillouin zone integrations used a $\Gamma$-centered Monkhorst–Pack $k$-point grid with spacing of 0.020 \AA$^{-1}$. All calculations included spin polarization.

\subsection{Moment tensor potential}

At its core, MTP represents the local atomic environment through moment tensors $M_{\mu,\nu}$, defined as
\begin{equation}
    M_{\mu,\nu}(\mathbf{n_i}) = \sum f_\mu(|\mathbf{r}_{ij}|, z_i, z_j) \, \mathbf{r}_{ij} \otimes \ldots \otimes \mathbf{r}_{ij}
\end{equation}
where $\mathbf{n}_i$ denotes the neighborhood of atom $i$, comprising the relative positions $\mathbf{r}_{ij}$ and types $z_j$ of neighboring atoms. The position vector $\mathbf{r}_{ij}$ points from atom $i$ to atom $j$, while $z_i$ and $z_j$ represent the atomic species of atom $i$ and its neighbor, respectively. Radial dependencies are captured by the functions $f_\mu$, whereas angular information is encoded via the tensor product ($\otimes$) of multiple $\mathbf{r}_{ij}$ vectors, resulting in a rank-$\nu$ tensor. The cutoff radius $r_c$ defines the maximum spatial range of interactions, while the maximum level of basis expansion ($\text{lev}_{\text{max}}$, an even integer typically between 2 and 28) controls the model's expressiveness by regulating the number of tensorial basis functions. During training, the objective function assigned different weights to each target quantity, with $w_\mathrm{e} = 1$ for energies, $w_\mathrm{f} = 0.01$ for forces, and $w_\mathrm{s} = 0.001$ for stresses, thereby prioritizing energy accuracy while still incorporating force and stress information to improve transferability. This weighting strategy ensured that the potential could accurately capture both local atomic environments and global energy trends. A 90:10 split between training and validation datasets was used to optimize fitting quality while monitoring overfitting. All MTP fitting, validation, and deployment were carried out using the MLIP package~\cite{novikov2020mlip} within the LAMMPS framework~\cite{thompson2022lammps}. Postprocessing, trajectory analysis, and visualization of atomic configurations were performed using OVITO~\cite{stukowski2009visualization}.

Following the recent methodology used for developing a Ti-Al-V potential within the MTP framework~\cite{nitol2025moment}, a similar approach was adopted in this work. A level-22 MTP basis was selected after comparison with level-18 models, yielding the lowest validation errors. The minimal cutoff radius, $r_{\text{min}}$, was initially set to 2.0~\AA\ and was adaptively updated during training, allowing the potential to dynamically capture short-range interactions critical for accurately modeling highly stressed atomic configurations. To optimize the maximum cutoff radius, $r_{\text{max}}$ was varied from 5.0 to 6.0~\AA\ in 0.1~\AA\ increments. For each combination of basis level and $r_{\text{max}}$, 20 models were trained with randomized initializations, resulting in a total of 400 models (2 levels $\times$ 10 cutoff values $\times$ 20 random seeds). Model selection was based on the lowest validation errors for both unary elastic constants and force predictions on quaternary 
datasets.

\section*{Data availability}
The potential, training database, and sample calculations are available on the author's GitLab page:
\url{https://gitlab.com/mtp_potentials/cocrfeni}.

\section*{Acknowledgment}
S.X. is grateful for the ORAU Ralph E. Powe Junior Faculty Enhancement Award.

\clearpage
\bibliography{cocrfeni}
\end{document}